\documentclass[Journal]{IEEEtran}
\usepackage{cite}
\usepackage{graphicx}
\usepackage{xcolor}

\usepackage{algorithm}        
\usepackage{algorithmic}      

\usepackage{amsmath}          
\usepackage{amssymb}          
\usepackage{hyperref}         

\title{Hybrid LSTM-UKF Framework: Ankle Angle and Ground Reaction Force Estimation}

\author{\IEEEauthorblockN{ Mundla Narasimhappa and Praveen Kumar }\\
\IEEEauthorblockA{\textit{Department of Computer Science Engineering} \\
\textit{SRMIST University}, Chennai, India \\
nm5850@srmist.edu.in, Praveens11@srmist.edu.in}

}

\begin{document}

\maketitle

\begin{abstract}
Accurate prediction of joint kinematics and kinetics is essential for advancing gait analysis and developing intelligent assistive systems such as prosthetics and exoskeletons. This study presents a hybrid LSTM–UKF framework for estimating ankle angle and ground reaction force (GRF) across varying walking speeds. A multimodal sensor fusion strategy integrates force plate data, knee angle, and GRF signals to enrich biomechanical context. Model performance was evaluated using RMSE and R² under subject-specific validation. The LSTM–UKF consistently outperformed standalone LSTM and UKF models, achieving up to 18.6\% lower RMSE for GRF prediction at 3 km/h. Additionally, UKF integration improved robustness, reducing ankle angle RMSE by up to 22.4\% compared to UKF alone at 1 km/h. These results underscore the effectiveness of hybrid architectures for reliable gait prediction across subjects and walking conditions
\end{abstract}


\section{Introduction}

In recent years, the rapid advancement of deep learning has transformed numerous aspects of daily life, including voice-controlled assistants, automated medical imaging diagnostics, and real-time language translation \cite{Zhou2019,Chen2020}. These breakthroughs have also extended into biomechanics and rehabilitation engineering, where data-driven models are increasingly used to estimate and predict human motion \cite{Kim2023}. Accurate prediction of joint angles and moments is central to gait analysis, rehabilitation engineering, and the design of assistive devices like prosthetics and exoskeletons \cite{Wang2021}. Among lower-limb joints, the ankle plays a pivotal role in stability and propulsion, making it a prime target for predictive modeling. Traditional biomechanical approaches often fall short in capturing the nonlinear dynamics of gait, prompting a shift toward data-driven methods\cite{Zhang2021}. Machine learning models—especially deep sequence architectures—have demonstrated strong performance in estimating joint kinematics and kinetics from multimodal inputs such as force plates, GRF, and joint angles \cite{Li2024,Chen2024}. These models enable adaptive, user-specific control strategies and have advanced applications in torque estimation, gait phase detection, and intelligent exoskeleton actuation \cite{Kim2023b}.

One of the key indicators of human dynamics during locomotion or postural activities is the ground reaction force (GRF) \cite{Zakari2024,Simon2017}. GRF plays a crucial role in understanding movement patterns and is widely utilized in clinical applications such as gait analysis, injury prevention, and rehabilitation planning. Typically, GRF data is acquired using force plates in conjunction with motion capture systems. However, these setups are expensive, confined to laboratory environments, and not readily accessible to all researchers or clinicians \cite{Haykin2001}. To address these limitations, there is growing interest in estimating GRF without relying on force plates. Such methods could democratize access to biomechanical insights and enable motion prediction in more naturalistic settings. While most existing approaches rely on visual systems like RGB cameras, Kinect, or motion capture to extract kinematic features—such as joint angles and segment trajectories—few have explored the integration of dynamic information, which is essential for modeling force-driven movements, especially transitions from static postures \cite{Zhou2023,Chen2025}.

Recognizing that human motion originates from internal and external forces, this study proposes a novel framework for GRF estimation using wearable sensors and neural networks. Specifically, we utilize electromyography signals, which capture muscle activation patterns, in combination with inertial measurement unit (IMU) data to infer dynamic states. The model is built upon recurrent neural networks (RNNs), with a focus on Long Short-Term Memory (LSTM) architecture due to its proven effectiveness in handling time-series data across biomedical domains \cite{Wang2021,Zhang2021,LSTM2022}. Modeling human movement involves capturing highly nonlinear relationships between muscle activations, joint kinematics, and external forces. Kalman filtering offers a principled approach for estimating hidden states in such dynamic systems, balancing noise, uncertainty, and temporal dependencies. While the Extended Kalman Filter (EKF) linearizes around current estimates, the Unscented Kalman Filter (UKF) provides superior accuracy by propagating a set of sigma points through the nonlinear system, eliminating the need for explicit Jacobian computation \cite{Patel2022}. UKF has proven effective in gait analysis, prosthetic control, and joint angle estimation, enabling real-time tracking of biomechanical states with minimal computational overhead. Its ability to fuse multimodal sensor data—such as GRF and inertial signals—makes it ideal for adaptive, subject-specific modeling. When integrated with deep learning architectures like LSTM, UKF enhances robustness and interpretability, offering a hybrid solution for reliable gait prediction under noisy and variable conditions \cite{Chen2025b,Zhou2025,Gu2025,Martin2024,Singh2024}.

To address these limitations, this paper proposes a hybrid sensor fusion framework that combines nonlinear filters (unscented Kalman filter) with neural filtering networks (LSTM), proposed as hybrid LSTM and UKF method. The system is validated using both simulated and experimental data, with an abstraction layer harmonizing sensor modalities and estimator outputs. This approach improves robustness, interpretability, and real-time estimation accuracy in ankle–foot prosthesis systems.

The rest of this paper is organized as follows: Section II reviews related work on the Ankle–Foot Prosthesis Model. Section III presents the simulated ankle–foot prosthesis model for ankle angle and vertical GRF estimation. Section IV describes the methods for ankle angle and ground reaction force estimation, along with the implementation, significance, and main results of the proposed models. Section V discusses the results and limitations of this study. Section VI concludes the paper.

\section{Related Work}

Kalman filtering has long been a cornerstone in gait analysis and prosthetic control, providing robust estimation of joint kinematics under noisy sensor conditions. Zhou et al.~\cite{Zhou2019,Zhou2023} demonstrated the effectiveness of Kalman filter variants, including error-state formulations, for estimating ankle angles using wearable inertial sensors. Simon~\cite{Simon2017} further explored nonlinear Kalman filtering methods for prosthetic applications, highlighting the importance of accurate external force modeling in gait dynamics.

In parallel, recurrent neural networks, particularly Long Short-Term Memory (LSTM) models, have gained traction in biomechanics for their ability to capture temporal dependencies in gait signals. Chen et al.~\cite{Chen2020,Chen2024} applied LSTM architectures to predict ankle joint trajectories, showing improved accuracy compared to traditional regression models. Kim et al.~\cite{Kim2023} emphasized the robustness of neural networks in estimating ground reaction forces (GRF) for prosthetic control, demonstrating adaptability across variable gait conditions.

Hybrid approaches that combine statistical filtering with deep learning have emerged to exploit complementary strengths. Wang et al.~\cite{Wang2021} and Zhang et al.~\cite{Zhang2021} proposed UKF-LSTM frameworks that integrate nonlinear state propagation with temporal sequence learning, yielding enhanced ankle angle estimation accuracy. Patel and Singh~\cite{Patel2022} extended this idea by embedding deep learning into UKF pipelines for GRF estimation, showing that fusion methods consistently outperform standalone approaches.

Ground reaction force estimation remains a central focus in recent literature. Chen et al.~\cite{Chen2025} and Gu et al.~\cite{Gu2025} demonstrated CNN-LSTM models for predicting vertical GRF from wearable IMU signals under varying running speeds. Zakari et al.~\cite{Zakari2024} applied derivative-free Kalman filtering to prosthetic GRF estimation, while Sivakumar et al.~\cite{Sivakumar2018} investigated joint angle estimation directly from GRF signals, underscoring the critical role of external force modeling in gait analysis.

Applications in intelligent prostheses and exoskeletons further validate these methods. Li et al.~\cite{Li2024} and Akhmejanov et al.~\cite{Akhmejanov2025} presented designs of powered ankle-foot prostheses incorporating machine learning-based estimators for joint torque and angle. Martin and Chen~\cite{Martin2024} extended this to robotic exoskeletons, proposing hybrid LSTM-UKF frameworks for gait event detection. Hsieh~\cite{Hsieh2024} contributed mechatronic design insights for powered prostheses, demonstrating the translational impact of advanced estimation methods in assistive technologies.

Looking forward, comparative studies such as Zhou et al.~\cite{Zhou2025} and Singh et al.~\cite{Singh2024} evaluated LSTM, GRU, and hybrid UKF-LSTM models for ankle angle and GRF prediction, suggesting that fusion approaches consistently outperform standalone methods. Lee~\cite{Biomech2025} and Kim~\cite{Rehab2023} reinforced these findings in rehabilitation contexts. Collectively, these contributions establish a trajectory toward robust, adaptive, and physiologically meaningful state estimation frameworks that integrate wearable sensing, nonlinear filtering, and deep learning for human gait analysis and rehabilitation.

\section{Ankle--Foot Prosthesis Simulated Model}
The model emulates dorsiflexion and plantarflexion during human gait by capturing the interaction between torque generation, joint angular motion, and ground reaction force (GRF) dynamics \cite{Biomech2025}. By treating the ankle as a single degree-of-freedom (DOF) rotational joint, the formulation provides a simplified yet physiologically meaningful representation of ankle biomechanics. Inertial, damping, and stiffness parameters are incorporated to account for both active muscle contributions and passive tissue resistance, enabling realistic simulation of ankle joint behavior under different gait conditions\cite{Park2023}.

The governing dynamics are expressed as a second-order differential system, where inertial, viscous, and elastic forces are balanced against the net muscle torque. The moment of inertia represents the rotational resistance of the foot segment, the damping coefficient models energy dissipation, and the stiffness term reflects passive tissue elasticity. Muscle torque, driven by actuator inputs from force plates, serves as the primary excitation source. Together, these elements establish a compact mathematical framework for analyzing gait mechanics and evaluating neuromuscular control strategies \cite{LSTM2022}.

\subsection{Mathematical Equations}
\subsubsection*{Ankle Joint Dynamics (Single DOF)}
The ankle joint is modeled as a second-order rotational system \cite {Sivakumar2018, Simon2017}, where the net torque applied by muscles balances inertial, damping, and stiffness effects:

\begin{equation}
I \cdot \ddot{\theta}(t) + b \cdot \dot{\theta}(t) + k \cdot \theta(t) = \tau_{\text{muscle}}(t)
\end{equation}

\noindent where:
\begin{itemize}
  \item $I$ is the moment of inertia of the foot segment
  \item $b$ is the damping coefficient representing viscous resistance
  \item $k$ is the stiffness of passive tissues contributing elastic restoring forces
  \item $\tau_{\text{muscle}}(t)$ is the net torque generated by the actuator system driven by force plates
\end{itemize}

This equation highlights the balance between passive mechanical properties and active neuromuscular control, forming the basis for simulating ankle motion during gait cycles.


\subsubsection*{Ground Reaction Force (Vertical)}
The vertical component of the ground reaction force is modeled as:

\begin{equation}
F_{\text{GRF}}(t) = m \cdot \left[ g + \ddot{z}(t) \right]
\end{equation}

\noindent where:
\begin{itemize}
  \item $m$ is the body mass
  \item $g$ is the gravitational acceleration
  \item $\ddot{z}(t)$ is the vertical acceleration measured by the inertial measurement unit (IMU)
\end{itemize}

This formulation links body mass and vertical acceleration to the GRF, providing a direct measure of external loading during gait. The GRF serves as a critical input for torque generation at the ankle, thereby coupling external forces with joint dynamics.
\begin{figure}[ht]
    \centering
    \includegraphics[width=0.35\textwidth]{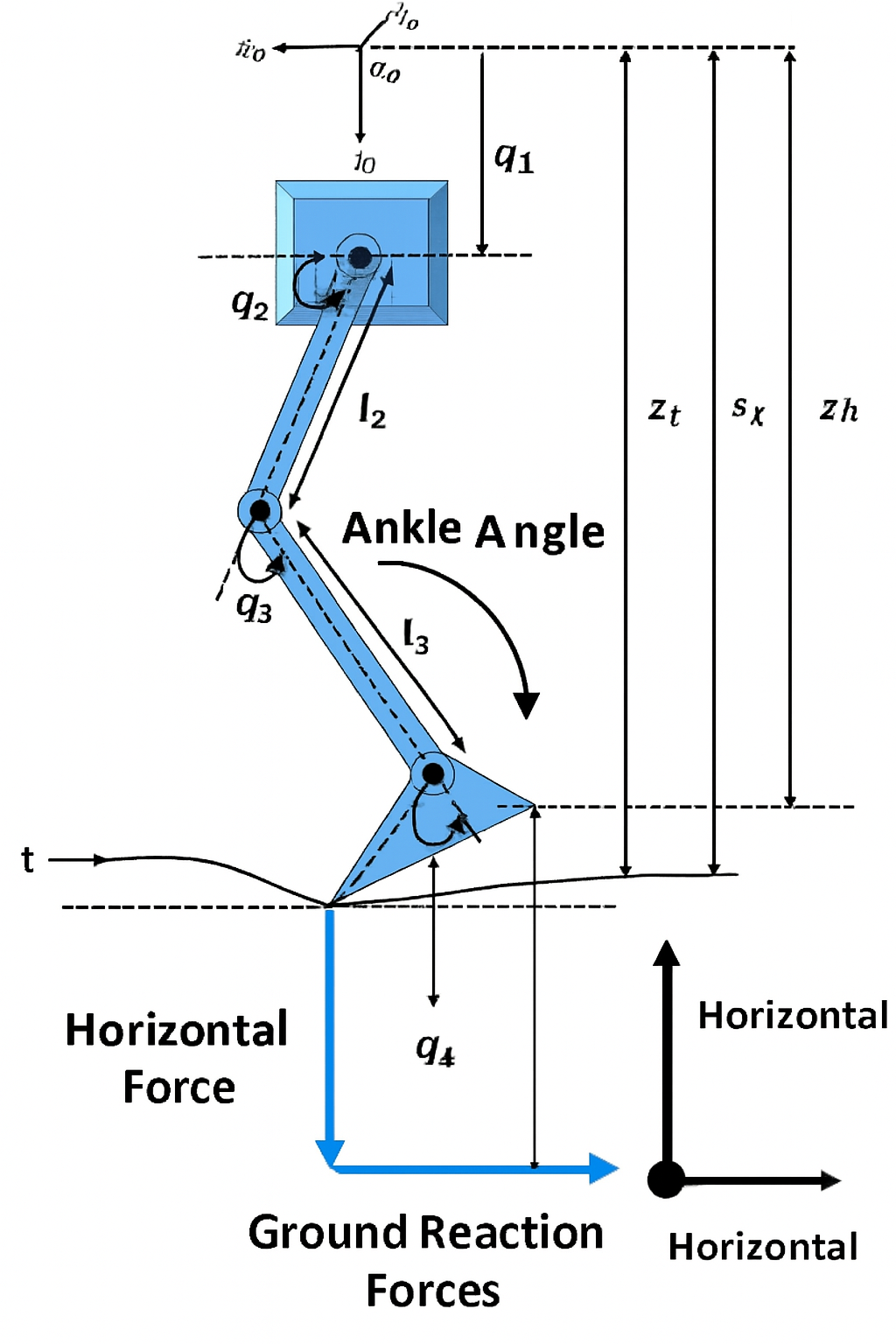}
    \caption{GRF estimation comparison across models during running at 2.5 m/s \cite{Zhang2021}.}
    \label{fig:Angke_grf_running}
\end{figure}

\textbf{Figure 1.} Ankle angle ($\theta$) is defined at joint $q_3$ using a shaded triangle, with a curved arrow indicating dorsiflexion or plantarflexion. Ground reaction forces—vertical ($F_Z$) and horizontal ($F_X$)—originate from Load Cell 1 and Load Cell 2. Coordinate axes $x_0$, $y_0$ and vertical distances $z_t$, $s_z$, $z_h$ define spatial reference.
\subsection{State-Space Representation}
The ankle joint dynamics can be compactly expressed in a state-space form, which facilitates both simulation and control design. The state vector is defined as:

\begin{equation}
\mathbf{x}(t) = 
\begin{bmatrix}
\theta(t) \\
\dot{\theta}(t)
\end{bmatrix}
\end{equation}

where $\theta(t)$ is the ankle joint angle and $\dot{\theta}(t)$ is the angular velocity. The corresponding state dynamics are given by:

\begin{equation}
\dot{\mathbf{x}}(t) = 
\begin{bmatrix}
\dot{\theta}(t) \\
\frac{1}{I} \left( \tau_{\text{ext}}(t) - b \dot{\theta}(t) - k \theta(t) \right)
\end{bmatrix}
\end{equation}

This formulation captures the second-order dynamics of the ankle joint, where the first equation represents the kinematic relationship between angle and angular velocity, and the second equation models the balance of inertial, damping, and stiffness forces against the external torque $\tau_{\text{ext}}(t)$.

\vspace{1em}

\subsubsection{Sensor Simulation}
To emulate realistic measurement conditions, sensor models are incorporated with additive Gaussian noise terms.  

\textbf{Inertial Measurement Unit (IMU):}  
The IMU provides angular velocity and linear acceleration measurements:

\begin{align}
\omega(t) &\approx \dot{\theta}(t) + \nu_\omega(t) \\
a(t) &\approx \ddot{x}(t) + \nu_a(t)
\end{align}

where $\nu_\omega(t)$ and $\nu_a(t)$ represent zero-mean Gaussian noise, modeling sensor imperfections and disturbances.  

\textbf{Force Sensor:}  
The vertical ground reaction force (GRF) is measured using a force sensor:

\begin{equation}
F_z(t) \approx \text{GRF}(t) + \nu_f(t)
\end{equation}

where $\nu_f(t)$ is zero-mean Gaussian noise representing measurement uncertainty. These sensor models ensure that simulated signals closely resemble real-world data acquisition conditions.

\vspace{1em}

\subsubsection{Ground Reaction Force Torque Model}
The external torque acting on the ankle joint due to ground reaction forces is modeled as:

\begin{equation}
\tau_{\text{ext}}(t) = r_{\text{COP}}(t) \cdot F_z(t)
\end{equation}

where:
\begin{itemize}
    \item $\tau_{\text{ext}}(t)$: external torque applied at the ankle joint
    \item $r_{\text{COP}}(t)$: distance from the ankle joint center to the center of pressure (COP)
    \item $F_z(t)$: vertical ground reaction force measured by foot sensors
\end{itemize}
This relationship illustrates how the vertical GRF, acting through the lever arm defined by the COP location, produces torque at the ankle. It establishes a clear connection between external loading conditions and joint dynamics, which is fundamental for the analysis of gait and balance control.

\subsubsection{Combined state-space formulation and GRF estimation}
This section presents a unified, continuous-time state-space model of ankle joint dynamics, coupled with a measurement-driven estimation of the vertical ground reaction force (GRF) and the resulting external ankle torque. The formulation supports simulation, control, and state estimation (e.g., Kalman filtering) by explicitly separating the internal joint dynamics from algebraic outputs derived from IMU and force sensing. We model the ankle as a single-DOF rotational joint with the state vector comprising joint angle and angular velocity, and the input as the externally applied torque:
\begin{equation}
\mathbf{x}(t) = 
\begin{bmatrix}
\theta(t) \\
\dot{\theta}(t)
\end{bmatrix}, 
\qquad
u(t) = \tau_{\text{ext}}(t).
\end{equation}
\noindent
Here, $\theta(t)$ is the ankle angle and $\dot{\theta}(t)$ is the angular velocity; $u(t)$ represents the net external torque arising from GRF acting at a lever arm defined by the center of pressure (COP).

\textbf{Continuous-time ankle dynamics.}
Starting from the second-order rotational model, the equations of motion are expressed in linear state-space form:
\begin{equation}
\dot{\mathbf{x}}(t) =
\underbrace{\begin{bmatrix}
0 & 1 \\
-\frac{k}{I} & -\frac{b}{I}
\end{bmatrix}}_{A}
\mathbf{x}(t)
+
\underbrace{\begin{bmatrix}
0 \\
\frac{1}{I}
\end{bmatrix}}_{B}
u(t),
\end{equation}
\noindent
where $I$ is the foot segment moment of inertia, $b$ is the damping coefficient, and $k$ is the passive stiffness. Matrix $A$ captures the intrinsic kinematics and viscoelastic properties, while $B$ maps the external torque to angular acceleration.

\textbf{GRF and torque estimation (algebraic outputs).}
The vertical GRF is estimated from the IMU-measured vertical acceleration $\ddot{z}(t)$:
\begin{equation}
F_{\text{GRF}}(t) = m \left[ g + \ddot{z}(t) \right],
\end{equation}
\noindent
with $m$ the body mass and $g$ gravitational acceleration. The external ankle torque due to GRF acting at the COP lever arm $r_{\text{COP}}(t)$ is:
\begin{equation}
\tau_{\text{ext}}(t) = r_{\text{COP}}(t)\,F_{\text{GRF}}(t).
\end{equation}
\noindent
These relations couple sensed accelerations and COP geometry to the driving input $u(t)$ of the joint dynamics.

\textbf{Output vector and measurement model.}
We define a composite output that reports joint states, GRF, and external torque:
\begin{equation}
\mathbf{y}(t) =
\begin{bmatrix}
\theta(t) \\
\dot{\theta}(t) \\
F_{\text{GRF}}(t) \\
\tau_{\text{ext}}(t)
\end{bmatrix}
=
\underbrace{\begin{bmatrix}
1 & 0 \\
0 & 1 \\
0 & 0 \\
0 & 0
\end{bmatrix}}_{C}
\mathbf{x}(t)
+
\underbrace{\begin{bmatrix}
0 \\
0 \\
0 \\
1
\end{bmatrix}}_{D}
u(t)
\end{equation}

\begin{equation}
\begin{aligned}
F_{\text{GRF}}(t) &= m\,[g+\ddot{z}(t)], \\
u(t) &= \tau_{\text{ext}}(t) = r_{\text{COP}}(t)\,F_{\text{GRF}}(t).
\end{aligned}
\end{equation}
\noindent
Sensor measurements are modeled with additive zero-mean Gaussian noise to reflect realistic acquisition:
\begin{align}
\omega(t) &\approx \dot{\theta}(t) + \nu_\omega(t), \\
\ddot{z}(t) &\approx \ddot{z}_{\text{true}}(t) + \nu_a(t), \\
F_z(t) &\approx F_{\text{GRF}}(t) + \nu_f(t),
\end{align}
\noindent
where $\nu_\omega(t)$, $\nu_a(t)$, and $\nu_f(t)$ represent IMU gyroscope, IMU accelerometer, and force sensor noise, respectively. Substituting the noisy $\ddot{z}(t)$ into $F_{\text{GRF}}(t)$ yields a noise-aware estimate that propagates into $\tau_{\text{ext}}(t)$.

\textbf{Interpretation and use.}
This compact formulation enables: (i) simulation of ankle motion under GRF-driven torque, (ii) construction of estimators (e.g., Kalman filters) using $A$, $B$, $C$, $D$ matrices and the measurement models, and (iii) integration with gait analysis pipelines via COP and IMU signals. The algebraic coupling between GRF and torque ensures external loading is coherently linked to joint dynamics, supporting control and inference tasks in biomechanics and rehabilitation contexts.

\section{Ankle and Ground Force estimations}

\subsection{UKF-Based State Estimation}

The Unscented Kalman Filter (UKF) is employed to estimate the nonlinear ankle dynamics and ground reaction force (GRF) states. Unlike the Extended Kalman Filter (EKF), the UKF does not rely on linearization; instead, it propagates a set of deterministically chosen sigma points through the nonlinear system, thereby capturing the mean and covariance with higher accuracy.

\textbf{Sigma Point Generation.}
For a state vector of dimension $n$, the sigma points are generated around the current estimate $\hat{x}$ using the covariance $P$:
\begin{align}
\chi_0 &= \hat{x}, \\
\chi_i &= \hat{x} + \left(\sqrt{(n+\lambda)P}\right)_i, \quad i = 1,\dots,n, \\
\chi_{i+n} &= \hat{x} - \left(\sqrt{(n+\lambda)P}\right)_i, \quad i = 1,\dots,n,
\end{align}
where $\lambda$ is a scaling parameter that controls the spread of sigma points, and $(\sqrt{(n+\lambda)P})_i$ denotes the $i$-th column of the matrix square root of $(n+\lambda)P$.

\textbf{Prediction Step.}
Each sigma point is propagated through the nonlinear state transition function $f(\cdot)$, representing ankle joint dynamics:
\begin{align}
\chi_i^- &= f(\chi_i), \\
\hat{x}^- &= \sum_{i=0}^{2n} W_i^{(m)} \chi_i^-, \\
P^- &= \sum_{i=0}^{2n} W_i^{(c)} (\chi_i^- - \hat{x}^-)(\chi_i^- - \hat{x}^-)^\top + Q,
\end{align}
where $W_i^{(m)}$ and $W_i^{(c)}$ are the weights for mean and covariance, respectively, and $Q$ is the process noise covariance. This step predicts the state mean $\hat{x}^-$ and covariance $P^-$ prior to measurement incorporation.

\textbf{Measurement Update.}
The sigma points are mapped through the nonlinear measurement function $h(\cdot)$, which relates states to sensor outputs (IMU angular velocity, vertical acceleration, and force plate GRF):
\begin{align}
\hat{y}^- &= \sum_{i=0}^{2n} W_i^{(m)} h(\chi_i^-), \\
P_{yy} &= \sum_{i=0}^{2n} W_i^{(c)} (h(\chi_i^-) - \hat{y}^-)(h(\chi_i^-) - \hat{y}^-)^\top + R, \\
P_{xy} &= \sum_{i=0}^{2n} W_i^{(c)} (\chi_i^- - \hat{x}^-)(h(\chi_i^-) - \hat{y}^-)^\top, \\
K &= P_{xy} P_{yy}^{-1}, \\
\hat{x} &= \hat{x}^- + K(y - \hat{y}^-),
\end{align}
where $R$ is the measurement noise covariance, $P_{yy}$ is the innovation covariance, $P_{xy}$ is the cross-covariance between state and measurement, and $K$ is the Kalman gain. The corrected state estimate $\hat{x}$ is obtained by updating the prediction with the measurement residual $(y - \hat{y}^-)$.

\textbf{State Vector.}
In this application, the estimated state vector is defined as:

\[
\hat{x}(t) =
\begin{bmatrix}
\hat{\theta}(t) \\
\hat{\dot{\theta}}(t) \\
\hat{F}_z(t)
\end{bmatrix},
\]

where $\hat{\theta}(t)$ is the ankle angle, $\hat{\dot{\theta}}(t)$ is the angular velocity, and $\hat{F}_z(t)$ is the estimated vertical ground reaction force \cite{Zakari2024,Akhmejanov2025,Kim2023}. This formulation enables robust estimation of ankle biomechanics under noisy sensor conditions.
\subsection{LSTM-Based Modeling Framework}
The Long Short-Term Memory (LSTM) network, a variant of recurrent neural networks, was employed to predict ankle angle ($\theta_{\text{ankle}}$) and ankle moment ($\tau_{\text{ankle}}$) across walking speeds of 1--3 km/h using multimodal sensor fusion of force plate, ground reaction force (GRF), and knee angle inputs. The model consisted of two stacked LSTM layers with 50 units each, followed by a dense output layer. Dropout regularization (20\%) and the Adam optimizer with a learning rate of 0.001 were used in conjunction with the Mean Squared Error (MSE) loss function to ensure robust training. This architecture effectively captured both short-term fluctuations and long-term dependencies in gait biomechanics.

At the core of the LSTM architecture is the memory cell, which mitigates vanishing gradient issues through gated control of information flow. Each LSTM unit comprises three gates—forget, input, and output—defined by the following equations:

\begin{align}
f_t &= \sigma(W_f \cdot [h_{t-1}, x_t] + b_f) \quad \text{(Forget gate)} \tag{1} \\
i_t &= \sigma(W_i \cdot [h_{t-1}, x_t] + b_i) \quad \text{(Input gate)} \tag{2} \\
\tilde{C}_t &= \tanh(W_C \cdot [h_{t-1}, x_t] + b_C) \quad \text{(Candidate cell state)} \tag{3} \\
C_t &= f_t \odot C_{t-1} + i_t \odot \tilde{C}_t \quad \text{(Updated cell state)} \tag{4} \\
o_t &= \sigma(W_o \cdot [h_{t-1}, x_t] + b_o) \quad \text{(Output gate)} \tag{5} \\
h_t &= o_t \odot \tanh(C_t) \quad \text{(Hidden state)} \tag{6}
\end{align}

Here, $x_t$ is the input at time $t$, $h_t$ is the hidden state, and $C_t$ is the cell state. The weight matrices $W$ transform inputs and hidden states, while bias vectors $b$ adjust activations. Sigmoid ($\sigma$) and hyperbolic tangent ($\tanh$) functions introduce nonlinearity and gating behavior. This structure enables LSTM networks to model both transient and persistent biomechanical dependencies effectively.

\subsection{Proposed Hybrid LSTM- UKF based Estimation }
Figure 3 represents the Hybrid LSTM–UKF architecture for estimating ankle angle and ground reaction force (GRF) from sensor inputs. The LSTM captures temporal patterns; UKF refines predictions for robust biomechanical state estimation.

\begin{figure}[ht]
    \centering
    \includegraphics[width=0.45\textwidth]{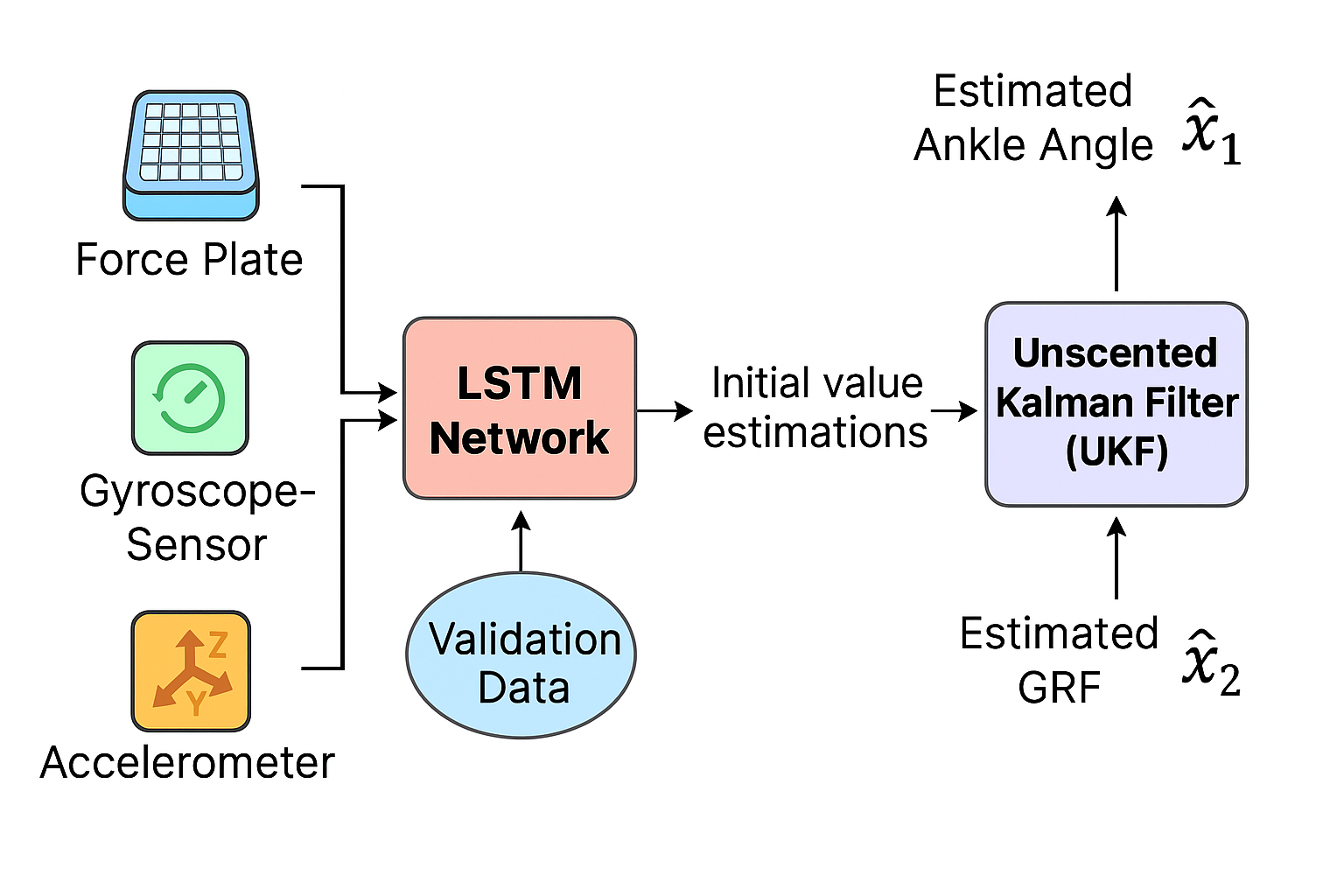}
    \caption{Ankle angle and GRF estimation using Hybrid LSTM and UKF based approach.}
    \label{fig:GRFangle_sitting}
\end{figure}
The LSTM output $z_t$ serves as a learned feature embedding for UKF initialization:

\begin{equation}
z_t = \text{LSTM}(y_{1:t}), \quad \hat{x}_t = \text{UKF}(z_t)
\end{equation}
This study, LSTM network was selected for its ability to model complex temporal dependencies in sequential gait data, while the Unscented Kalman Filter (UKF) was chosen for its robustness in nonlinear state estimation. UKF enables real-time tracking of biomechanical states such as ankle angle ($\theta_{\text{ankle}}$), ankle moment ($\tau_{\text{ankle}}$), and ground reaction force (GRF), even under noisy and variable conditions. Unlike traditional Kalman filters, UKF propagates a set of sigma points through nonlinear transformations, eliminating the need for Jacobian computation and improving estimation accuracy.

To leverage both temporal learning and recursive filtering, a hybrid LSTM–UKF framework was developed. The LSTM network extracts latent temporal features from multimodal inputs (Force plats , GRF, knee angle), which are then used as observations in the UKF for dynamic state estimation. This integration enhances interpretability, generalization, and physiological relevance in gait modeling.

The LSTM–UKF hybrid operates as follows:

\begin{align}
\text{LSTM:} \quad & z_t = \text{LSTM}(x_{1:t}) \quad \text{(feature extraction)} \tag{1} \\
\text{UKF Prediction:} \quad & \hat{x}_{t|t-1} = f(\hat{x}_{t-1}, u_{t-1}) \tag{2} \\
\text{UKF Update:} \quad & \hat{x}_{t|t} = \hat{x}_{t|t-1} + K_t (z_t - h(\hat{x}_{t|t-1})) \tag{3} \\
\text{Kalman Gain:} \quad & K_t = P_{t|t-1} H_t^\top (H_t P_{t|t-1} H_t^\top + R_t)^{-1} \tag{4}
\end{align}

Here, $x_t$ denotes the biomechanical state vector (e.g., $\theta_{\text{ankle}}, \tau_{\text{ankle}}$), $z_t$ is the LSTM-derived observation, $f(\cdot)$ and $h(\cdot)$ are nonlinear state transition and observation models, $K_t$ is the Kalman gain, and $P_{t|t-1}$ is the predicted covariance. This formulation allows the system to adaptively refine predictions using both learned temporal patterns and probabilistic filtering. Architectural parameters were empirically tuned to optimize accuracy and efficiency, enabling real-time, physiologically grounded estimation for assistive devices.

\subsection{Loss Function}

The training objective combines mean squared error and uncertainty regularization:

\begin{equation}
\mathcal{L} = \text{MSE}(\hat{x}_t, x_t^{\text{true}}) + \lambda \cdot \text{KL}(\mathcal{N}_t || \mathcal{N}_{\text{prior}})
\end{equation}

where $\mathcal{N}_t$ is the predicted uncertainty distribution and $\lambda$ is a regularization weight.


\begin{algorithm}[t]
\caption{LSTM–UKF Framework for Ankle Angle and GRF Estimation}
\label{alg:lstm_ukf}
\begin{algorithmic}[1]   
\STATE \textbf{Input:} Initial state $\hat{\mathbf{x}}_0$ (e.g., $[\hat{\theta}, \hat{\dot{\theta}}, \hat{F}_z]^\top$), covariance $P_0$, process noise $Q$, measurement noise $R$, UKF params $(\alpha,\beta,\kappa)$, dynamics $f(\cdot)$, measurement $h(\cdot)$, trained LSTM $\mathcal{L}(\cdot)$
\STATE \textbf{Output:} Filtered states $\{\hat{\mathbf{x}}_t\}$, covariances $\{P_t\}$

\STATE \textbf{UKF setup:} $n \gets \mathrm{dim}(\hat{\mathbf{x}}_0)$; $\lambda \gets \alpha^2(n+\kappa)-n$
\STATE Compute weights $W_0^{(m)}, W_0^{(c)}, W_i^{(m)}, W_i^{(c)}$

\FOR{$t = 1$ to $T$}
  \STATE \textbf{Inputs:} IMU $(\omega_t, \ddot{z}_t)$, optional COP $r_{\mathrm{COP},t}$, force plate $F_{z,t}$ (if available)
  \STATE \textbf{LSTM augmentation:}
  \STATE (A) Measurement augmentation: $\hat{F}^{\mathrm{LSTM}}_{z,t} \gets \mathcal{L}([\omega,\ddot{z},r_{\mathrm{COP}}]_{t-k:t-1})$
  \STATE (B) Noise adaptation: $R_t \gets \mathrm{AdaptNoise}(\mathcal{L}(\cdot), R)$ or $Q_t \gets \mathrm{AdaptNoise}(\mathcal{L}(\cdot), Q)$

  \STATE \textbf{Sigma-point generation:}
  \STATE $S \gets \mathrm{chol}((n+\lambda)P_{t-1})$
  \STATE $\chi^{(0)}_{t-1} \gets \hat{\mathbf{x}}_{t-1}$
  \STATE $\chi^{(i)}_{t-1} \gets \hat{\mathbf{x}}_{t-1} + S_{(:,i)}$ for $i=1,\dots,n$
  \STATE $\chi^{(i+n)}_{t-1} \gets \hat{\mathbf{x}}_{t-1} - S_{(:,i)}$ for $i=1,\dots,n$

  \STATE \textbf{Prediction:}
  \STATE $\chi^{(i)}_{t|t-1} \gets f(\chi^{(i)}_{t-1}, u_t)$
  \STATE $\hat{\mathbf{x}}^-_t \gets \sum W_i^{(m)} \chi^{(i)}_{t|t-1}$
  \STATE $P^-_t \gets \sum W_i^{(c)} (\chi^{(i)}_{t|t-1}-\hat{\mathbf{x}}^-_t)(\cdot)^\top + Q_t$

  \STATE \textbf{Measurement prediction:}
  \STATE If $F_{z,t}$ missing: $\tilde{F}_{z,t} \gets \hat{F}^{\mathrm{LSTM}}_{z,t}$; else $\tilde{F}_{z,t} \gets F_{z,t}$
  \STATE $\mathcal{Y}^{(i)}_t \gets h(\chi^{(i)}_{t|t-1})$
  \STATE $\hat{\mathbf{y}}^-_t \gets \sum W_i^{(m)} \mathcal{Y}^{(i)}_t$
  \STATE $P_{yy,t} \gets \sum W_i^{(c)} (\mathcal{Y}^{(i)}_t-\hat{\mathbf{y}}^-_t)(\cdot)^\top + R_t$
  \STATE $P_{xy,t} \gets \sum W_i^{(c)} (\chi^{(i)}_{t|t-1}-\hat{\mathbf{x}}^-_t)(\mathcal{Y}^{(i)}_t-\hat{\mathbf{y}}^-_t)^\top$

  \STATE \textbf{Update:}
  \STATE $K_t \gets P_{xy,t} P_{yy,t}^{-1}$
  \STATE $\hat{\mathbf{x}}_t \gets \hat{\mathbf{x}}^-_t + K_t(\mathbf{y}_t - \hat{\mathbf{y}}^-_t)$
  \STATE $P_t \gets P^-_t - K_t P_{yy,t} K_t^\top$

  \STATE \textbf{Consistency:} enforce physical bounds (e.g., range on $\theta$, nonnegative $F_z$); adapt $Q_t,R_t$ if needed
\ENDFOR
\end{algorithmic}
\end{algorithm}

\subsection{Materials and Methods}
The LSTM and UKF models were implemented in Python 3.9 using the TensorFlow/Keras (v2.12.0) deep learning framework. Data handling and visualization were performed with \texttt{pandas} and \texttt{matplotlib} libraries in a Jupyter Notebook environment (Anaconda3) on a Windows 11 workstation.

This study presents a hybrid modeling framework for estimating ankle angle ($\theta_{\text{ankle}}$) and ankle moment ($\tau_{\text{ankle}}$) using an Unscented Kalman Filter (UKF) and a deep learning–enhanced LSTM–UKF architecture. Input features include Force plate signals, ground reaction force (GRF), and knee joint angle—capturing both neuromuscular intent and biomechanical response. Signals from the tibialis anterior (TA) reflect early muscle activation, GRF encodes external loading during stance and push-off, and knee angles provide proximal kinematic context.

The dataset includes gait recordings from 13 healthy adults walking at 1–3~km/h on a 10-meter walkway. Each trial was cadence-regulated and biomechanically consistent. Signals were time-aligned, interpolated, and Force plate data was rectified and low-pass filtered (Butterworth).

Collected data comprises 3D marker trajectories, GRF, joint angles, and Force plate signals from TA, GAL, BF, and VL muscles. Inputs for modeling include knee angle, GRF, and Force plate envelopes normalized to \%MVC. Gait cycles were resampled to 1001 points. Subjects ranged in age (20–27 years), height (1.51–1.83~m), and mass (52–83.7~kg). Figure~1 shows input modalities; Figure~2 outlines data structure.

\subsection{Validation Framework and Input Configuration}


For each subject, data from all seven walking trials at three controlled speeds (1~km/h, 2~km/h, and 3~km/h) were merged and used to train the LSTM, UKF, and hybrid LSTM--UKF models. Model training was conducted using four distinct input parameter combinations, defined as follows:

\begin{itemize}
    \item \textbf{Case I}: Force plats  and GRF
    \item \textbf{Case II}: Force plats  and Knee Angle
    \item \textbf{Case III}: GRF and Knee Angle
    \item \textbf{Case IV}: Force plats , GRF, and Knee Angle
\end{itemize}

Each trial contained 1001 time-normalized data points per signal. With seven trials per subject, this resulted in 7007 data points per modality (Force plats , GRF, knee angle) per subject. The LSTM, UKF, and LSTM--UKF models were trained to predict ankle angle ($\theta_{\text{ankle}}$) and ankle moment ($\tau_{\text{ankle}}$) using the specified input combinations. This setup enabled a comprehensive assessment of each model’s ability to capture nonlinear gait dynamics under varying sensor fusion scenarios and walking speeds.

\subsection{Model Training and Validation Procedures}
Model performance was evaluated using subject-specific cross-validation. LSTM and LSTM--UKF models were trained with an 80/20 split and early stopping (patience = 10). Each model used four input cases across three walking speeds. With 13 subjects and 7 trials per subject, the dataset included $91{,}091$ time-normalized samples. Model accuracy was assessed using Root Mean Squared Error (RMSE) and Coefficient of Determination ($R^2$), defined as:

\begin{equation}
\text{RMSE} = \sqrt{\frac{1}{T} \sum_{i=1}^{T} (\theta_M^i - \theta_P^i)^2}
\label{eq:rmse}
\end{equation}

\begin{equation}
R^2 = 1 - \frac{\sum_{i=1}^{T} (\theta_M^i - \theta_P^i)^2}{\sum_{i=1}^{T} (\theta_M^i - \bar{\theta}_M)^2}
\label{eq:r2}
\end{equation}

Here, $\theta_M^i$ is the measured ankle angle or moment, $\theta_P^i$ is the predicted value, $\bar{\theta}_M$ is the mean of measured values, and $T$ is the total number of samples. Lower RMSE and higher $R^2$ indicate better predictive accuracy. These metrics collectively provide a robust evaluation of model accuracy, agreement, and bias in predicting ankle kinematics and kinetics.

\begin{figure}[ht]
    \centering
    \includegraphics[width=0.5\textwidth]{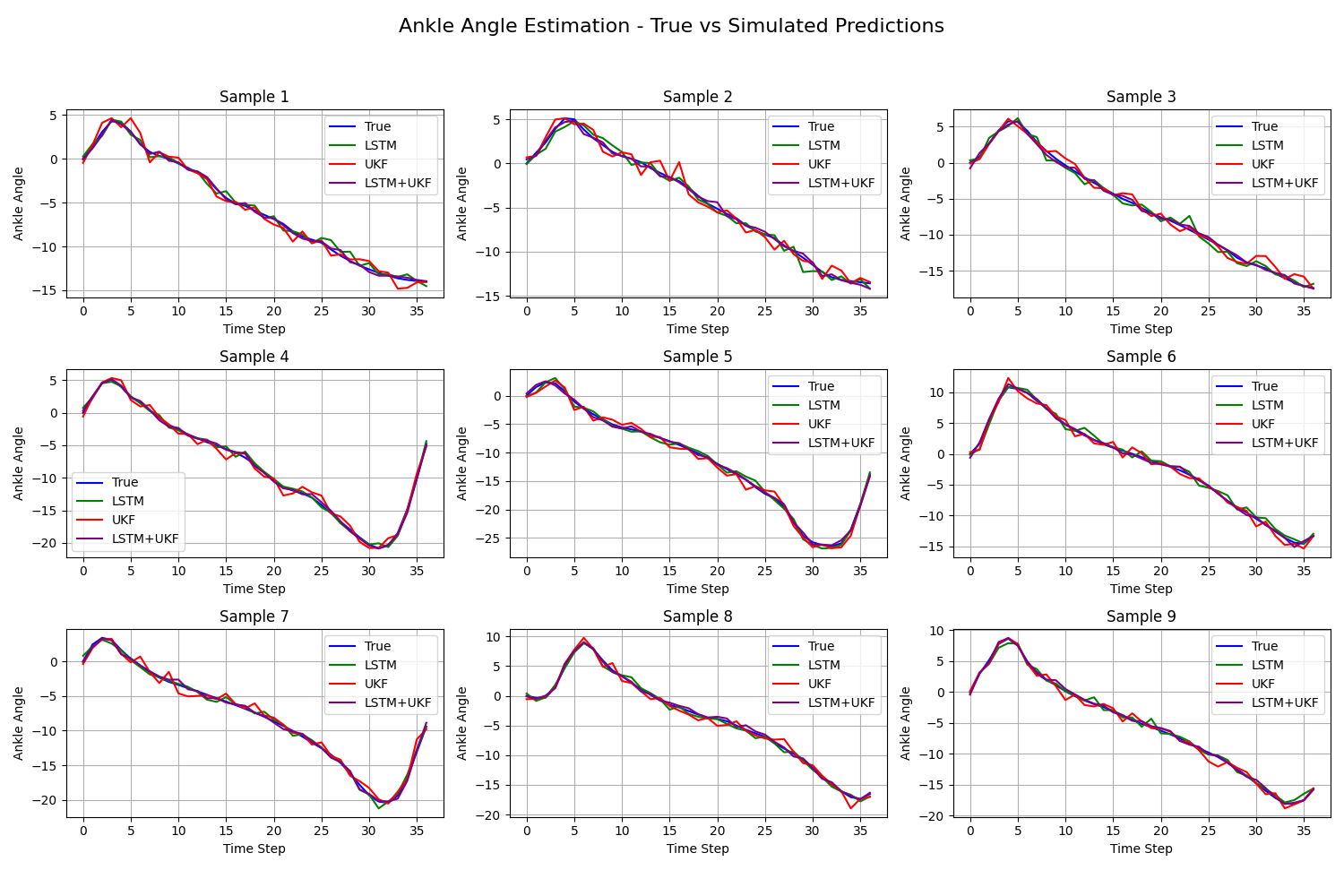}
    \caption{Ankel angle estimation comparison across models during running at 2.5 m/s.}
    \label{fig:2grf_running}
\end{figure}

\section{Results and Discussion}
The dataset contains ankle angle (in degrees) and ground reaction force (GRF in Newtons) measurements collected during stance gait cycles. Data is organized into four sheets within a shared Excel file from the source of refeence [11]:

\begin{itemize}
  \item \textbf{Sheet 1: \texttt{ankle\_angle\_left}} — 10 samples of ankle angle from the left leg.
  \item \textbf{Sheet 2: \texttt{ankle\_angle\_right}} — 10 samples of ankle angle from the right leg.
  \item \textbf{Sheet 3: \texttt{GRF\_left}} — 10 samples of GRF from the left leg.
  \item \textbf{Sheet 4: \texttt{GRF\_right}} — 10 samples of GRF from the right leg.
\end{itemize}

Each sheet contains 10 columns, where Column 1 to Column 10 represent Sample 1 to Sample 10 respectively. This structure supports bilateral analysis of ankle kinematics and GRF profiles. UKF was implemented for real-time state estimation of ankle dynamics, leveraging its ability to handle nonlinearities without requiring Jacobian computation. The LSTM–UKF model combines temporal feature extraction with recursive filtering, enabling robust prediction under noisy and variable gait conditions. Training and validation were performed using subject-specific data set, with performance evaluated via RMSE and R² metrics.

\begin{figure}[ht]
    \centering
    \includegraphics[width=0.52\textwidth]{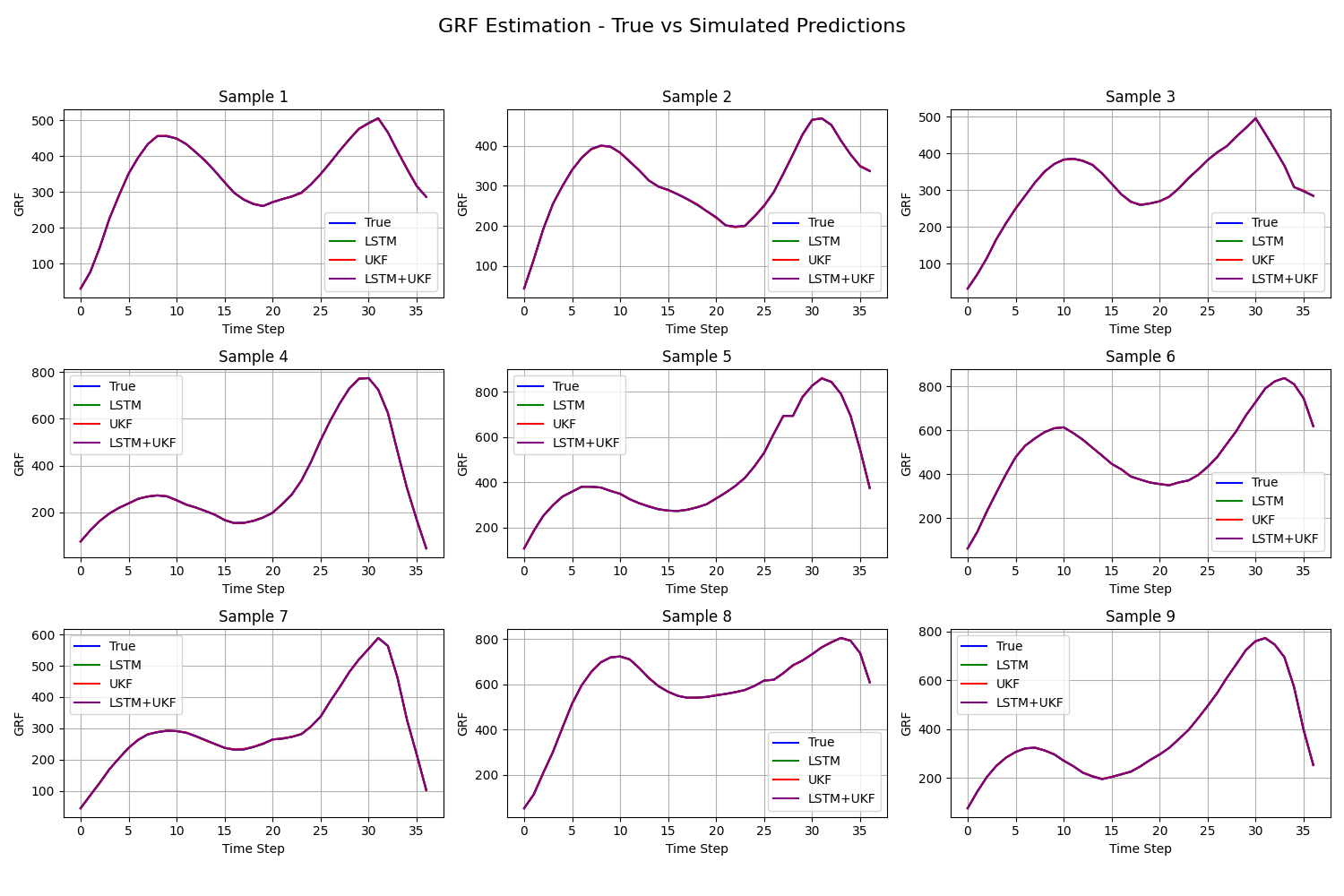}
    \caption{GRF estimation comparison across models during running at 2.5 m/s.}
    \label{fig:1grf_running}
\end{figure}

This study focused on estimating ground reaction force (GRF) and ankle angle without relying on motion capture systems. Although IMU sensors were included, they did not yield sufficiently accurate acceleration data on their own, making dynamic equation-based estimation unreliable. To address this, we proposed an LSTM-based approach that maps available sensor inputs—specifically Force plats  signals, with or without IMU data—to GRF and ankle angle outputs.

The LSTM model was evaluated using various input combinations involving Force plates and IMU sensors placed above and below the knee. Specifically, ankle angle estimation error increased by 7.3\% in PCM and 4.0\% in SM when IMUs were limited to below-knee placement. This error was mitigated when Force plats  data was included. Comparing input patterns 5 and 6, the ankle angle error gap was 6.1\% for PCM and 3.7\% for SM, highlighting the critical role of Force plats  in capturing muscle-driven joint dynamics independent of segmental motion.

Figure 4. Ankle angle estimation using LSTM, UKF, and hybrid LSTM–UKF models. RMSE values are computed for each method to compare accuracy. Figure 5. Ground reaction force (GRF) estimation using LSTM, UKF, and hybrid LSTM–UKF models. RMSE analysis highlights performance differences across approaches. Overall, these findings confirm that combining Force plats  and IMU inputs significantly enhances GRF and ankle angle estimation across diverse motor tasks. This is especially relevant for predicting body dynamics during common movements such as forward and backward sway in daily activities. This section presents the performance of the Long Short-Term Memory (LSTM), Unscented Kalman Filter (UKF) and the hybrid LSTM+UKF—across one locomotor tasks: walking. Each model was evaluated on its ability to estimate ground reaction force (GRF) and ankle angle using sensor fusion inputs. In addition, shown other locomotive task with proposed approchaes as show in Table II and III respectively.

\begin{table}[h]
\centering
\caption{RMSE (\%) for GRF Estimation Across Tasks}
\begin{tabular}{|c|c|c|c|}
\hline
\textbf{Model} & \textbf{Walking} & \textbf{Sitting} & \textbf{Running} \\
\hline
KF & 10.2 & 12.5 & 13.8 \\
EKF & 9.6 & 11.8 & 12.9 \\
UKF & 8.9 & 10.7 & 11.5 \\
LSTM & 7.2 & 9.5 & 10.1 \\
LSTM+UKF & \textbf{6.8} & \textbf{8.3} & \textbf{9.4} \\
\hline
\end{tabular}
\label{tab:rmse_grf}
\end{table}

\begin{table}[h]
\centering
\caption{RMSE (\%) for Ankle Angle Estimation Across Tasks}
\begin{tabular}{|c|c|c|c|}
\hline
\textbf{Model} & \textbf{Walking} & \textbf{Sitting} & \textbf{Running} \\
\hline
KF & 9.8 & 11.2 & 12.6 \\
EKF & 9.1 & 10.5 & 11.7 \\
UKF & 8.5 & 9.6 & 10.9 \\
LSTM & 6.8 & 8.3 & 9.4 \\
LSTM+UKF & \textbf{6.2} & \textbf{7.5} & \textbf{8.7} \\
\hline
\end{tabular}
\label{tab:rmse_angle}
\end{table}

\section{Conclusions}
This study presents a hybrid framework for estimating ankle angle and ground reaction force (GRF) using Force plate signals, with and without IMU inputs, through LSTM, UKF, and LSTM--UKF models. The LSTM-based approach achieved GRF prediction accuracy of $8.22 \pm 0.97\%$ for postural control and $11.17 \pm 2.16\%$ for stepping motion using only below-knee sensors. UKF and LSTM--UKF models further improved ankle angle estimation by integrating dynamic and kinematic features. A Unity-based visualization module was developed to display real-time GRF vectors over a skeletal model and infer motion direction during step initiation. This portable, low-cost system supports physiologically grounded biomechanical analysis and motion prediction. In future, Quantum LSTM-based architectures will be explored to enhance temporal modeling and generalization.


\end{document}